\documentclass[a4paper]{article}

\usepackage{INTERSPEECH2020}
\usepackage{multirow, multicol}
\usepackage{arydshln}
\usepackage{url,times,booktabs}
\usepackage{mathtools}
\usepackage{cite}

\title{Segment Aggregation for short utterances speaker verification using raw waveforms}
\name{Seung-bin Kim, Jee-weon Jung, Hye-jin Shim, Ju-ho Kim, Ha-Jin Yu$^\dag$}
\address{
  School of Computer Science, University of Seoul, Republic of Korea}
\email{kimho1wq@naver.com, jeewon.leo.jung@gmail.com, shimhz6.6@gmail.com, wngh1187@naver.com, hjyu@uos.ac.kr}

\begin{document}

\maketitle
\begin{abstract}
Most studies on speaker verification systems focus on long-duration utterances, which are composed of sufficient phonetic information. 
However, the performances of these systems are known to degrade when short-duration utterances are inputted due to the lack of phonetic information as compared to the long utterances.
In this paper, we propose a method that compensates for the performance degradation of speaker verification for short utterances, referred to as \textit{``segment aggregation''}.
The proposed method adopts an ensemble-based design to improve the stability and accuracy of speaker verification systems. 
The proposed method segments an input utterance into several short utterances and then aggregates the segment embeddings extracted from the segmented inputs to compose a speaker embedding.
Then, this method simultaneously trains the segment embeddings and the aggregated speaker embedding.
In addition, we also modified the teacher-student learning method for the proposed method.
Experimental results on different input duration using the VoxCeleb1 test set demonstrate that the proposed technique improves speaker verification performance by about 45.37\% relatively compared to the baseline system with 1-second test utterance condition.

\end{abstract}

\section{Introduction}
Many research studies have been carried out to improve the performance of speaker verification using Deep neural networks (DNNs), which have demonstrated the state-of-the-art performance \cite{d-vector, snyder2018x, Voxceleb2, safari2019self}.
A speaker verification system refers to a system that verifies the authenticity of a speaker using speech characteristics.
The information extracted by a speaker verification system may include speaker-specific information, etc., and the amount of such information may affect the performance of the system.
Such information can easily be exploited when the duration of the speech is long and most speaker verification studies have been conducted using long utterances.

However, compared to long utterances, short utterances may not contain all the speech characteristics that can be obtained from voice.
In this case, uncertainty arises when extracting an utterance-level feature because there is less speaker-specific information used to train the system.
Therefore, the performance of the system is reported to greatly degrade when short utterances are input, and this is due to the increased uncertainty in the short utterances  \cite{poddar2018shortutt, poddar2019shortutt}.
To solve this problem, research studies have to focus on designing speaker verification systems that are capable of authenticating both short and long utterances.

Ensemble technique is widely used to obtain better prediction performance than the case of using learning algorithms separately \cite{Deeplearningbook, Residual, hu2018squeeze}.
Bootstrap aggregating (bagging) technique is an ensemble learning method that averages multiple estimates to reduce the variance of an estimate \cite{bagging}.
Given a training dataset, bagging creates several small-sized training sets by uniformly sampling from the dataset, and then various weak predictors are generated by training with each small-sized training set.
The results of each generated bagging predictor are combined to make a final decision—that is to produce a final predictor with high performance.

Inspired from the bagging technique, we propose a novel method to improve the performance for short-duration utterances in speaker verification.
Our objective is to ensure that the performance of the system is not affected by the length of input utterances.
Our method makes the system robust to short utterances by training with short utterance segments and long utterances by using ensemble aggregation of segment embeddings extracted from the segmented utterances.
Unlike the bagging method that creates various weak predictors, the proposed method develops a single predictor that generates multiple internal representations.
In this paper, we refer to this method as segment aggregation (SA).

The SA produces several short utterances by segmenting an input utterance into short utterance segments, and then these short utterances are simultaneously input into a shared network in parallel.
The network produces several segment embeddings from the input segmented utterances, and the segment embeddings are aggregated into a single speaker embedding.
The aggregated speaker embedding is connected to the output layer that performs speaker identification.
To reduce the variance between the segment embeddings that are extracted from segmented utterances, we simultaneously train the speaker verification system with these segment embeddings and the aggregated speaker embedding.
In addition, we train the system to maximize the cosine similarity of the aggregated speaker embedding and the original speaker embedding of baseline system extracted from a long utterance to improve the performance for long utterances.

The rest of this paper is organized as follows. Section 2 describes related works with a baseline system and speaker verification systems for short utterances. Section 3 introduces our proposed method and Section 4 describes our proposed method with teacher-student learning. Section 5 shows experiments and results and conclusions are presented in Section 6.

\section{Related Works}
\subsection{Raw waveform based DNN}
\begin{table}[t] 
  \caption{Architecture of the modified RawNet. Batch normalization and LeakyReLU are applied before the convolution layer in the residual block, except for the first block \cite{he2016identity}.}
  \centering
  \label{tab:DNN_arch}
  \begin{tabular}{lcc}
  \toprule
  \textbf{Layer} & \textbf{Input: raw wave ($T \times 1$)} & \textbf{Output size}\\
   
  \midrule
  & Conv(3,3,128) & \multirow{3}{*}{$T/3 \times 128$}\\
  Stride-conv& BN & \\
  & LeakyReLU & \\
  \midrule
  Res block & 
    
    $\left \{
      \begin{tabular}{c}
      Conv(3,1,128)\\
      Conv(3,1,128)\\
      MaxPool(3)\\
      \end{tabular}
    \right \}$ 
    $\times$2
  & $T/27 \times 128$\\

  \midrule
  Res block & 
    
    $\left \{
      \begin{tabular}{c}
      Conv(3,1,256)\\
      Conv(3,1,256)\\
      MaxPool(3)\\
      \end{tabular}
    \right \}$ 
    $\times$4
  & $T/2187 \times 256$\\
  \midrule
  GRU & GRU(1024) & $1024$\\
  \midrule
  Speaker & \multirow{2}{*}{FC(1024)} & \multirow{2}{*}{$1024$}\\
  embedding & & \\
  \midrule
  Output & FC(6112) & $6112$\\
  \bottomrule
  \end{tabular}
\end{table}

\begin{table}[h]
  \renewcommand\thetable{2}
  \caption{Comparison of the original system and the modified version of RawNet. Performances are reported using EER on the original VoxCeleb1 test set.}
  \label{tab:baseline}
  \centering
  \begin{tabular}{lcc}
    \toprule
    System & Trained on & EER (\%)\\
    \midrule
    \# 1-RawNet \cite{jung2019rawnet} & VoxCeleb 1 & 4.80 \\
    \# 2-Baseline & VoxCeleb 2 & \textbf{3.50} \\
    \bottomrule
  \end{tabular}
\end{table}

Many recent studies have used less processed features for training DNN based speaker embedding extractor, and many research studies have reported that DNNs based on direct modeling of raw waveforms have several advantages over DNNs modeled with conventional acoustic features  \cite{hajibabaei2018unified, jung2018avoiding22, ravanelli2019sincnet}.
The reason for using raw waveforms is that as the size of data increases, the probability that DNNs extract the information needed for each task from raw waveforms increases, and performance can be improved \cite{jung2019rawnet, muckenhirn2018towards, ravanelli2018speaker}.
In addition, by using raw waveforms, the exploration of various hyper-parameters to extract acoustic features is not required.
For this reason, we adopt RawNet \cite{jung2019rawnet}, which takes raw waveforms as input, as the speaker embedding extractor.

We used the modified version of the RawNet architecture described in Table 1 as the baseline system.
Table \ref{tab:baseline} describes the performance of the original RawNet trained on the VoxCeleb1 dataset, referred to as system \# 1, and our modified version of RawNet trained on VoxCeleb2 dataset, referred to as system \# 2.
Results from our experiments show that our baseline system improve performance over the original system, leading to a relative error reduction (RER) of 27.1\%.
The proposed method is applied to the system \# 2.

\subsection{Speaker verification systems for short utterances}

Various methods have been proposed to improve the performance of speaker verification systems for short utterances.
\cite{jung2019short} proposed a short utterance compensation framework in speaker verification that maximizes the cosine similarity of two speaker embeddings extracted from long and short utterances.
\cite{xie2019aggregation} proposed an utterance-level aggregation method with a NetVLAD or GhostVLAD layer in the wild scenario.
This layer is adopted for the application of a self-attentive pooling method with a learnable dictionary encoding.
\cite{hajavi2019short} proposed a time-distributed voting (TDV) aggregation system for short-segment speaker recognition.
This system extracts as much information as possible from a single utterance and then selects useful information.
Similar to \cite{hajavi2019short}, we extract useful information from a single utterance, but train a system using intuitive ensemble technique without using any pooling method, such as self-attentive pooling and TDV.

\section{Segment Aggregation}
\begin{figure*}[t]
  \centering
  \includegraphics[width=\linewidth]{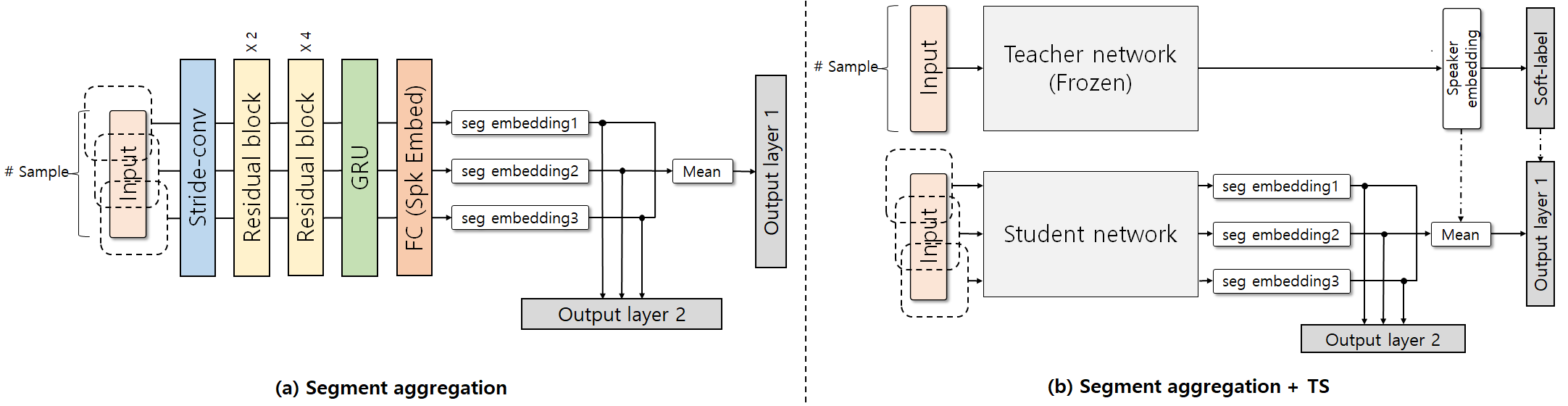}
  \caption{Proposed methods to improve the performance on short utterances. \textbf{(a)}: The segment aggregation system. The segment embeddings extracted from segmented utterances are aggregated using the average function. Segment embeddings and the aggregated speaker embedding are used simultaneously for training the output layers using categorical cross-entropy for speaker identification. \textbf{(b)}: The segment aggregation system with teacher-student learning. The student network utilizes speaker embeddings and soft-labels created by the teacher network for training.}
  \label{fig:overall}
\end{figure*}

One of the well-known techniques to compensate for the poor performance of speaker verification systems for short utterances is to train the systems using short utterances in the training phase.
However, this above-stated technique increases systems’ robustness for short utterances but degrades the systems’ overall performance for long utterances.\footnote{As a result of internal experiments, performance for long utterances deteriorated when short utterances are used for training.}
This result seems to have occurred because the network is overfitting for short utterances with strong uncertainty, and accordingly, the information is excessively omitted to consider for uncertainty even when a long utterance is entered.
To solve this problem, we segment long duration utterances into several short utterance segments and train a network using the short utterance segments in parallel.
The segment embeddings extracted from the segmented input are element-wisely averaged to compose a speaker embedding, and this speaker embedding is connected to the output layer of the network that performs the speaker identification.
We refer to this technique as segment aggregation (SA) and the illustration of the overall system is depicted in Figure 1-(a).

Let $\boldsymbol{x}$ be an input utterance of any speaker, $\boldsymbol{x} \in \mathbb{R}^{F}$, where $F$ refers to the number of the samples in the training phase (length of sequence).
Given an input utterance $\boldsymbol{x}$ of any speaker, a network segments the input utterance into $K$ short utterances $\boldsymbol{x}_{k}\in\mathbb{R}^{C}$, $k=1,...,K$, where $C$ is the length of each segment.
The network simultaneously extracts segment embeddings from each short utterance segment and subsequently aggregates the segment embeddings into a speaker embedding.
The speaker embedding is derived as follows:
\begin{equation}
    \boldsymbol{e} = \frac{1}{K}\sum_{k}^{K}e_{k}
\end{equation}
where $e$ denotes an aggregated speaker embedding of an utterance, $K$ refers to the number of segments in an utterance and $e_{k}$ denotes a segment embedding extracted from a segment $\boldsymbol{x}_{k}$.
Lastly, the speaker embedding is connected to an output layer which is trained for speaker identification using categorical cross-entropy (CCE) objective function.

For example, using SA technique, a segment length is first set. 
When the segment length is set to 2s with an overlap of 1s with a mini-batch size of 6s, five segment utterances will be created by each input utterance, and accordingly, five segment embeddings are extracted by inputting these segment utterances in parallel into the network.

This method optimizes aggregated speaker embeddings averaged from segment embeddings.
However, there is a possibility that the variance of the segment embeddings increases.
This is because the method optimizes for speaker embeddings and does not optimizes each segment embedding directly, and the average value can be constant even if the variance of the segment embeddings is large.
Therefore, we further propose a method to reduce the variance of the segment embeddings. 

To increase the accuracy of segment embeddings, we simultaneously train the segment embeddings and the aggregated speaker embedding in separate output layers.
Finally, the objective function $Loss_{sa}$ for SA technique is defined as follows: 
\begin{equation}
    Loss_{sa} = Loss_{e} + W\sum_{k}^{K}Loss_{e_{k}}
\end{equation}
where $Loss_{e}$ denotes CCE for an output layer that receives an aggregated speaker embedding, $W$ denotes a weight for $Loss_{e_{k}}$, and $Loss_{e_{k}}$ denotes CCE for an output layer that receives a segment embedding.

\section{Teacher-student learning}

The teacher-student (TS) learning method was first proposed for model compression and is being used in a variety of fields \cite{li2014learning, li2018developing, kim2018bridgenets, jung2019short}.
\cite{jung2019short} uses two networks of the same architecture and size.
A teacher network (TN) that is pre-trained with long utterances transfers useful information such as soft-label and speaker embedding to a student network (SN).
Then, the SN is trained with short utterances to yield the correct answer similar to the received speaker embedding and soft-label.

The existing TS learning method for short utterances is designed to maximize the cosine similarity of two speaker embeddings extracted from long and short utterances thereby compensating for the performance for short utterances.
Similarly, we make the speaker embedding aggregated from short utterance segment embeddings with high uncertainty to be as close as possible to the original speaker embedding extracted from long utterances.
Figure 1-(b) depicts our system that uses the teacher-student learning.

We create a RawNet as a TN and a RawNet with the proposed SA technique as an SN in order to adopt TS learning architecture, and input utterances of the same duration into the two networks.
Let $e_T(x)$ be a speaker embedding extracted from TN and $e_S(x)$ be the aggregated speaker embedding extracted from SN, where $x$ refers a long input utterance.
The objective function $Loss_{ts}$ for the modified TS learning is defined as follows: 
  \begin{equation}
    \begin{aligned}
    Loss_{ts} = \sum_{j}^{J}Cos(e_T(x_j), e_S(x_j))\\ - \sum_{j}^{J}\sum_{i}^{I}P_T(s_i|x_j)log(P_S(s_i|x_j))
    \end{aligned}
  \end{equation}
where $Cos( , )$ denotes cosine similarity for two speaker embeddings, $i$ and $j$ refer to the speaker and utterance indices, and $P_T(s_i|x)$ and $P_S(s_i|x)$ are probabilities for any speaker $s_i$ for TN and SN respectively.
We add $Loss_{ts}$ and $Loss_{sa}$ for applying the TS learning method to our proposed system.

\section{Experiments and results}
We implemented the system with the PyTorch library \cite{paszke2019PyTorch}.
Code for experiments in this paper is freely available.\footnote{https://github.com/kimho1wq/SegmentAggregation}

\subsection{Dataset}
We used the VoxCeleb2 dataset \cite{Voxceleb2} in the training phase and VoxCeleb1 dataset \cite{Voxceleb} in the validation and test phase.
VoxCeleb1 contains approximately 330 hours of audio recordings from 1251 speakers for text-independent scenarios.
VoxCeleb2 has emerged as an extended version of the VoxCeleb1 dataset and contains over a million utterances from 6112 speakers.
We used all the utterances of VoxCeleb2 for training and utterances of 1211 speakers of VoxCeleb1 as validation data, and utterances of 40 speakers of VoxCeleb1 as test data.
\begin{table*}[t]
  \renewcommand\thetable{4}
  \caption{Performance comparison of state-of-the-art speaker verification systems that adopted methods to improve performance for short utterances and are trained on VoxCeleb2 dataset. Performances is reported EER on the original VoxCeleb1 test set.}
  \centering
  \label{tab:sota}
  \begin{tabular}{l c c c c c c c c}
  \toprule
  & \multirow{2}{*}{Model} & \multirow{2}{*}{Method} & \multirow{2}{*}{Input Feature}
  & 3 sec, & 2 sec, & 1 sec, & Full-length, \\
  & & & & EER (\%) & EER (\%) & EER (\%) & EER (\%) \\
  
  \midrule 
  \midrule
  Xie \textit{et. al.} \cite{xie2019aggregation} & Thin ResNet34 & GhostVlad & Spectrogram & 5.47 & 7.69 & 13.20 & 3.22\\
  Jung \textit{et. al.} \cite{jung2019short} & RawNet & TS & Raw waveform & 4.91 & 7.12 & 14.40 & 3.49 \\
  \textbf{Ours} & RawNet & SA & Raw waveform & 5.38 & 7.41 & 12.82 & 3.63 \\
  \textbf{Ours} & RawNet & SA + TS & Raw waveform & \textbf{4.59} & \textbf{6.05} & \textbf{11.15} & \textbf{3.15}\\
  \bottomrule
  \rule{0in}{1ex}
  \end{tabular}
\end{table*}

\begin{table}[h]
  \renewcommand\thetable{3}
  \caption{Results of our proposed system compared to the baseline with different duration. The segment length for applying SA technique is set to a fixed value or a random value. Performances is reported in EER.}
  \label{tab:ca_ts}
  \centering
  \renewcommand{\tabcolsep}{3mm}
  \begin{tabular}{lcccc}
    \toprule
    \multirow{2}{*}{System} & Segment & 3 sec, & 2 sec, & 1 sec,\\
     & length & EER & EER & EER \\
    
    \midrule
    Baseline & - & 6.64 & 8.93 & 20.41 \\
    \# 3-SA & 1 sec & 5.97 & 7.63 & 12.41 \\
    \# 4-SA & 2 sec & 5.49 & 7.38 & 14.46 \\
    \# 5-SA & 1-2 sec & 5.38 & 7.41 & 12.82 \\
    \midrule
    \# 6-SA + TS & 1 sec & 5.02 & 6.39 & \textbf{10.95} \\
    \# 7-SA + TS & 2 sec & 4.87 & 6.11 & 13.13 \\
    \# 8-SA + TS & 1-2 sec & 4.64 & 6.17 & 11.21 \\
    \# 9-SA + TS & 1-3 sec & \textbf{4.59} & \textbf{6.05} & 11.15 \\
    \bottomrule
  \end{tabular}
\end{table}

\subsection{Experimental configurations}
We input pre-emphasized raw waveforms into the network and configured the mini-batch for training by cropping the duration of input utterances to $59049$ samples ($\approx$ 3.69 s).
To evaluate the performances of the speaker verification systems on short utterances, we cropped the test utterances into different lengths of 1, 2 and 3 seconds—we set $16038$ samples to a length of 1s, $32076$ samples to a length of 2s, and $48114$ samples to a length of 3s.
When using the SA technique, we divided input utterances by overlapping about 10\% of the segment length.
An output of the last fully-connected layer is used as a segment embedding for using SA technique and the speaker embedding’s dimensionality is $1024$.

We used Leaky ReLU activation functions \cite{leaky} with a negative slope of $0.3$, AMSGrad optimizer \cite{reddi2019convergence} with a learning rate of $0.001$ and weight decay with $\lambda=1e^{-4}$.
We used categorical cross-entropy for all output layers.
We did not use any augmentation technique for training and test.

\subsection{Results analysis}

Table \ref{tab:ca_ts} shows the results of applying our proposed methods to the baseline system with different utterance duration.
System \# 3, 4, 6 and 7 use a fixed segment length, and the other systems use a different segment duration for each mini-batch in the training phase.
The result of the baseline system shows performances of system \# 2 with various lengths.
System \# 3, 4 and 5 are generated by applying the SA technique to the baseline system with varying segment lengths.
We set the weight of loss function $Loss_{e_{k}}$ to $0.2$ to give more weight to the loss function of the aggregated speaker embedding $Loss_{e}$.
Experimental results of these three systems confirmed the improved performance compared to the baseline with all test utterance conditions.
The system trained with fixed segment length shows improved performance on test sets with fixed segment lengths, whereas the system trained with different segment lengths showed improved average performance on test sets with varying lengths.
The last four rows in Table 3 describe the results of applying the TS learning method to the SA system.
To experiment with the application of TS learning method, we set the weight of loss function $Loss_{e_{k}}$ to $1.0$ because the loss function $Loss_{ts}$ for the teacher-student learning relatively reduces the weight of existing loss function $Loss_{e_{k}}$.
Results of these systems show that applying TS method to the SA system further improved the performance, especially when the segment length is randomly generated with a value between 1 to 3 seconds—the  average performance is most improved.

Table \ref{tab:sota} shows the performance comparison of state-of-the-art speaker verification systems that adopted different methods to improve performance for short utterances on the original VoxCeleb1 test set.
We couldn't directly compare the performance in \cite{xie2019aggregation} and \cite{hajavi2019short} because these studies report performances using self-curated trials and are not freely available.
However, the code of \cite{xie2019aggregation} is freely available, so using this code we retested their system on the original VoxCeleb1 test set and compared its performance.
Results show that our system using SA method (system \# 5) outperforms the performance of state-of-the-art systems when using 1-second test utterances with EER of 12.82\%.
The system adopting the SA and TS methods (system \# 9), which has the best average performance, outperforms for all length of test utterances than other start-of-the-art systems.
System \# 9 demonstrates an RER of 45.37\% compared to the modified RawNet and an RER of 22.57\% compared to the RawNet that applied TS learning method with 1-second test utterance condition.

\section{Conclusions}
In this paper, we propose a novel method to improve the performance of a speaker verification system when short-duration utterances are input. 
Our proposed method makes a system robust to short utterances by training the system with short utterance segments and long utterances by using ensemble aggregation of segment embeddings extracted from segmented utterances.
The method segments an input utterance into several shorter utterances and aggregates the segment embeddings extracted from the segmented utterances into a speaker embedding.
Also, the proposed method simultaneously trains multiple segment embeddings and the aggregated speaker embedding to reduce the variance between the segment embeddings.
In addition, we apply the teacher-student learning method to the proposed system to improve the performance of the aggregated speaker embedding.
We use the intuitive ensemble technique which divides the existing long utterance into several short utterances to achieve high robustness for short utterances.
Experimental results are reported using EERs with different input duration from the VoxCeleb1 test set.
Experimental results show that the system that applied our proposed method and the TS learning method has improved average performance for both long and short utterances of different duration.
Notably, the system showed an improved performance of around 45.37\% compared to the baseline system with a 1-second test utterance condition.
 
\section{Acknowledgement}
This work was supported by the Technology Innovation Program (10076583, Development of free-running speech recognition technologies for embedded robot system) funded by the Ministry of Trade, Industry \& Energy(MOTIE, Korea)

\newpage

\bibliographystyle{IEEEtran}
\bibliography{refs}
\end{document}